# SDU

University of Southern Denmark

**A Visualization Framework for Exploring Multi-Agent-Based Simulations: Case Study of an Electric Vehicle Home Charging Ecosystem**

Christensen, Kristoffer; Jørgensen, Bo Nørregaard; Ma, Zheng Grace

Go to publication entry in University of Southern Denmark's Research Portal

# A Visualization Framework for Exploring Multi-Agent-Based Simulations: Case Study of an Electric Vehicle Home Charging Ecosystem

Kristoffer Christensen[1]*[0000-0003-2417-338X], Bo Nørregaard Jørgensen[1][0000-0001-5678-6602] and Zheng Grace Ma[1]*[0000-0002-9134-1032]

[1] SDU Center for Energy Informatics, Maersk Mc-Kinney Moeller Institute, The Faculty of Engineering, University of Southern Denmark, Odense, Denmark

kric@mmmi.sdu.dk, bnj@mmmi.sdu.dk and zma@mmmi.sdu.dk

**Abstract.** Multi-agent-based simulations (MABS) of electric vehicle (EV) home charging ecosystems generate large, complex, and stochastic time-series datasets that capture interactions between households, grid infrastructure, and energy markets. These interactions can lead to unexpected system-level events, such as transformer overloads or consumer dissatisfaction, that are difficult to detect and explain through static post-processing. This paper presents a modular, Python-based dashboard framework—built using Dash by Plotly—that enables efficient, multi-level exploration and root-cause analysis of emergent behavior in MABS outputs. The system features three coordinated views (System Overview, System Analysis, and Consumer Analysis), each offering high-resolution visualizations such as time-series plots, spatial heatmaps, and agent-specific drill-down tools. A case study simulating full EV adoption with smart charging in a Danish residential network demonstrates how the dashboard supports rapid identification and contextual explanation of anomalies, including clustered transformer overloads and time-dependent charging failures. The framework facilitates actionable insight generation for researchers and distribution system operators, and its architecture is adaptable to other distributed energy resources and complex energy systems.

**Keywords:** Multi-agent simulation, data visualization, electric vehicles, dashboard, smart grid, emergent behavior.

## 1 Introduction

Agent-Based Modeling (ABM) is a computational approach that simulates the actions and interactions of autonomous agents to analyze complex systems and emergent phenomena [1]. By representing individual entities—such as households, vehicles, or market participants—with distinct behaviors and decision rules, ABM enables the study of emergent phenomena that arise from local interactions. In practice, these models are often implemented as Multi-Agent-Based Simulations (MABS), where large numbers of agents operate within dynamic environments over time. MABS has become a widely used methodology for modeling complex socio-technical systems, generating rich, high-resolution datasets that capture detailed temporal and spatial patterns across thousands of agents. While this granularity is essential for understanding emergent phenomena, it also introduces significant challenges: raw simulation outputs are typically stored in flat-file formats (e.g., CSV), resulting in data that is both voluminous and difficult to analyze without dedicated post-processing tools [2, 3]. Effectively interpreting these outputs requires advanced data processing pipelines and interactive visualization frameworks that allow users to explore patterns, trace anomalies, and compare scenarios across multiple analytical levels.

Within the energy domain, MABS are increasingly applied to simulate residential electricity networks—particularly in scenarios involving the widespread adoption of distributed energy resources such as electric vehicles (EVs), heat pumps, and solar photovoltaics. In these contexts, agent behaviors (e.g., charging, heating, or consumption) are governed by heterogeneous rules and stochastic inputs, leading to complex and often unexpected system-level outcomes [4, 5]. A particularly pressing challenge arises in understanding the impacts of EV home charging on low-voltage distribution grids, where synchronized charging behavior can overload critical infrastructure components such as 10/0.4 kV transformers [6]. Identifying the causes of such overload events requires not only aggregate system analysis but also the ability to drill down to individual agent behavior—examining charging schedules, departure times, battery states, and pricing signals in a temporally coherent manner.

In response, this paper introduces a modular, Python-based dashboard framework for the interactive exploration of MABS outputs, with a specific focus on EV charging simulations in residential distribution networks. The system addresses the analytical gap between large, high-resolution simulation datasets and the need for interpretable, decision-support tools by providing:

1. Automated transformation of raw outputs into optimized formats;
2. a multi-page dashboard with interactive visualizations (e.g., time-series, heatmaps, spatial charts); and
3. Coordinated drill-down capabilities for agent-level analysis and anomaly tracing.

The contributions of this work are threefold: (i) a generalizable dashboard architecture tailored to post-simulation analysis of MABS; (ii) a demonstration of its utility in diagnosing emergent behaviors in a full-year, high-frequency EV charging simulation; and (iii) methodological guidance for researchers applying visualization to complex simulation output data. As the use of MABS continues to expand in smart grid research, this work addresses a critical bottleneck: the lack of accessible, scalable, and user-friendly tools for extracting actionable insights from rich agent-level simulations.

## 2 Related work

MABS are increasingly used to study complex socio-technical systems, including energy transitions, EV charging ecosystems, and smart grid interactions. These simulations generate high-frequency, multivariate datasets that are often difficult to ana-lyze using static post-processing alone. Recent literature has emphasized the importance of advanced visualization and interactive tools to support analysis, communication, and decision-making in such contexts [2].

## 2.1 Visualization in Energy and Mobility ABMs

Visualization has long played a central role in helping researchers and stakeholders interpret ABM outputs. In energy and mobility domains, geospatial, temporal, and network-based displays are commonly used to illustrate emergent patterns at multiple scales. For example, in energy adoption studies, spatial heatmaps are used to show geographic clustering of distributed energy resources such as solar panels or EV chargers [3]. In mobility models, vehicle flows or traveler trajectories are often mapped in 2D or 3D to highlight congestion zones or behavioral trends [2, 7]. These approaches help link individual agent behaviors to system-level phenomena but often rely on preconfigured, static visualizations.

Several tools—commercial (e.g., AnyLogic) and open-source (e.g., GAMA, Repast)—support embedded GIS integration and agent mapping [8], enabling visual overlays of model outcomes on real-world geography. However, these tools are often tied to their simulation runtime environments, limiting post-hoc, drill-down exploration flexibility. More recent efforts have addressed this by developing specialized analysis and visualization frameworks that separate simulation and evaluation layers. For instance, AgentLens introduces a visual analytics system for exploring Large Language Model-based agent behaviors through temporal decomposition and traceable cause–effect mapping [9]; Similarly, AgentPy [10] offers interactive exploration of agent behaviors in Python environments using dynamic charts and live controls. These systems reflect a growing trend toward flexible, post-simulation visualization tools, as also pursued in this work.

## 2.2 Temporal and Interactive Visualization Approaches

Given the time-evolving nature of MABS outputs, time-series plots and animations are widely used to trace system dynamics such as energy demand, charging load, or emissions profiles [11]. Animated replays can highlight spatio-temporal behavior—e.g., EV charging peaks during nighttime hours or dissatisfaction events triggered by scheduling conflicts. However, raw animations often lack analytical precision, and static plots can obscure agent-level variability.

Recent work advocates for more interactive visualization frameworks capable of linking system-level metrics to individual agent actions. For instance, Grignard et al. [12] propose "Agent-Based Visualization", where visual elements are treated as agents in their own right, enabling immersive inspection in 3D environments. Similarly, the GAMA and Mesa platforms offer browser-based interfaces that support real-time charts, parameter manipulation, and filtering of agents [11].

Despite these advances, the literature reveals persistent gaps in post-simulation analysis tools for MABS, especially in energy domains with large agent populations and high temporal resolution. Many studies report only aggregate system Key Per-formance Indicators (KPIs) (e.g., total load, average emissions) with limited ability to diagnose specific emergent events, such as transformer overloads or charging failures. Moreover, few tools support drill-down analysis that traces a system-level anomaly to its underlying causes—such as the behavior of specific EVs, their state-of-charge, trip history, or charging strategy.

### 2.3 Need for Scalable, Multi-Level Exploration Tools

This paper addresses this gap by introducing a dashboard framework tailored for agent-level post-simulation analysis of EV-grid interactions. Unlike embedded visualization modules or static reporting tools, our approach supports synchronized exploration across system, transformer, and consumer layers. It enables researchers and distribution system operators (DSOs) to explain emergent, high-impact events by interrogating time-aligned data streams at multiple levels of granularity. This type of analysis—currently underrepresented in the literature—is essential for interpreting complex charging ecosystems and guiding future energy policy and infrastructure design.

Despite these advances, a gap remains in tools that support post-simulation exploration of high-frequency, stochastic energy data at multiple analytical levels. Many studies emphasize aggregate KPIs but lack support for tracing emergent phenomena to specific agent behaviors—e.g., transformer overloads caused by clustered EV charging. This paper addresses this gap through a modular dashboard tailored for synchronized, agent-level investigation.

## 3 Case Study: EV Home Charging

The dashboard framework is applied to a MABS of an EV home charging ecosystem in the Danish residential network of Strib, comprising 126 consumer nodes. The underlying model and simulation environment are based on the laxity-based aggregation strategy proposed in [6]. For this demonstration, the scenario assumes 100% EV adoption with smart charging behavior, where vehicles attempt to charge during periods of lowest electricity price. The simulation spans a full calendar year (2025) with a temporal resolution of one minute.

In such high-resolution, agent-driven simulations, emergent behaviors—such as transformer overloads, suboptimal charging, or consumer dissatisfaction—are common and often non-trivial to diagnose through static or aggregate analyses alone. Understanding the root causes of these phenomena requires layered insight across multiple dimensions of agent behavior. For example, explaining a specific overload event involves identifying which EVs were actively charging at the time, and further drilling down into each consumer's behavior, including their driving patterns (i.e., distance driven before and after the event), charging schedules, EV model specifications (battery capacity and charging power), state-of-charge (SoC) upon arrival, baseload consumption, and respective arrival and departure times. The dashboard enables precisely this kind of structured, drill-down investigation—turning what would otherwise be an opaque anomaly into an interpretable, data-driven narrative. By making such analysis intuitive and reproducible, the tool bridges the gap between raw simulation output and actionable insight.

The MABS model employed here is adopted from [6], which presents a laxity-based EV aggregation strategy aimed at minimizing charging costs while avoiding transformer overload. The model simulates individual households with varying base-load consumption, EV models, and trip profiles, and it incorporates a smart charging mechanism that shifts demand to low-price periods. The outputs from this simulation—covering variables such as charging load, SoC, arrival and departure times, and grid stress indicators—serve as the primary input to the dashboard framework described in this work. Readers are referred to [6] for detailed modeling assumptions and algorithmic descriptions.

## 4 Dashboard System Architecture

The presented dashboard framework is implemented using Dash by Plotly, a Python-based web application framework specifically designed for creating interactive, analytical dashboards. This choice was driven by several key considerations:

**Tight Integration with Python and Plotly.** Dash is natively built for Python, making it highly compatible with the broader Python ecosystem used for data processing and scientific computing (e.g., Pandas, NumPy, and PyArrow). Most critically, it leverages Plotly.js under the hood—a robust, open-source JavaScript graphing library that supports high-quality, publication-ready interactive visualizations. This ensures that the dashboard can seamlessly generate time-series plots, heatmaps, and geospatial charts with full interactivity, pan/zoom, and export support directly from Python without needing separate front-end code.

**Web-Native Architecture for Flexible Deployment.** Dash applications are rendered as HTML/CSS/JavaScript in the browser, meaning they are web-native by design. This architecture facilitates platform-independent deployment on local machines, intranets, or cloud services (e.g., Heroku, AWS, or internal servers). Because the entire interface is defined in Python, it simplifies development for teams that may not have front-end developers, while also enabling easy adaptation for broader access via browser-based tools.

**Modularity and Customization.** Dash supports a modular, callback-driven design, which is essential for building coordinated multi-view layouts (e.g., linking maps with time-series plots or enabling drill-down analysis from KPIs to agent-level charts). Furthermore, it supports plug-ins like Dash Bootstrap Components for more advanced layouts and styling, aligning with the needs of a professional-grade visualization tool.

Selecting the appropriate framework for building an interactive analytical dashboard is a critical design decision, particularly in simulation-heavy domains such as energy systems and smart grids. The dashboard must support not only visually compelling plots, but also the ability to handle large volumes of high-resolution, agent-level data and provide flexible, responsive interactivity for exploration and insight generation.

To contextualize the selection of Dash by Plotly, this section compares it against several other commonly used dashboard development tools—Power BI, Tableau, Streamlit, and Shiny—across a range of technical and practical dimensions. These include language environment, visualization capabilities, interactivity, scalability, deployment options, and fit for scientific simulation use cases.

A comparison overview can be found in Table 1 in the Appendix. The comparison illustrates that while tools like Power BI and Tableau excel in business intelligence applications, they are generally less suited to the demands of scientific simulations and domain-specific analytics. These platforms offer limited customization, are tightly coupled to proprietary ecosystems, and are optimized for tabular data rather than large-scale, time-resolved agent-based simulations.

Streamlit and Shiny offer more flexibility for technical users and are well-suited for prototyping and lightweight applications. However, their simplicity can be a limitation when building multi-page, highly coordinated dashboards with complex user interactions and visual hierarchies.

In contrast, Dash by Plotly provides the best alignment with the project requirements: it is Python-native, integrates directly with the scientific computing stack (e.g., Pandas, Plotly, NumPy), supports scalable deployment through web-native rendering (HTML/CSS/JavaScript), and offers extensive customization through callback-driven logic and modular components. Its interoperability with Plotly ensures high-quality, publication-ready visualizations that can be exported in SVG format for reporting or academic use. These features collectively make Dash an ideal choice for simulation practitioners who need fine-grained control over data exploration, agent-level insight, and visual storytelling within the context of energy systems modeling.

The dashboard's architecture follows visualization design principles proposed by Munzner [13], emphasizing domain-driven task abstraction and coordinated visual idioms aligned with multi-agent simulation data. To support this structure and aid understanding of the dashboard's structural organization, Fig. 1 presents a high-level architecture diagram of the system. It illustrates the transformation flow from raw simulation outputs through data processing components to the coordinated visualization views, highlighting the modular design and interaction between back-end scripts and front-end dashboard elements.

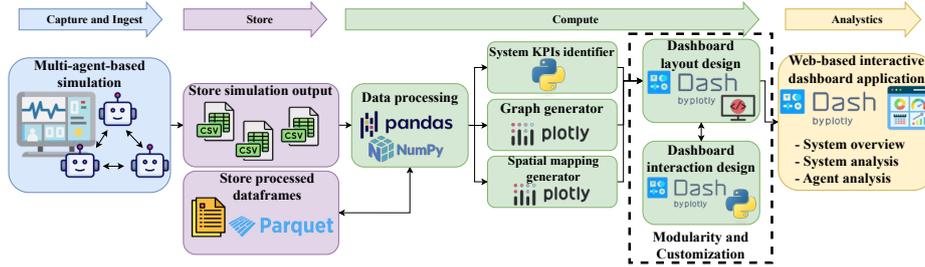

**Fig. 1.** System architecture of the dashboard framework.

### 4.1 Data Processing Pipeline

Simulation outputs are generated as CSV files, containing, among others, agent-level time series data for baseload consumption, EV consumption, EV battery State-of-Chart (SoC), $CO_2$ emissions, electricity prices, and network tariffs. The data pipeline consists of multiple Python modules imported into a central app.py file:

- **data_processing.py** reading data files, formatting time columns, calculating system metrics used as KPIs in the dashboard, and includes a suite of utility functions for preparing specific visualizations.
- **dashboard_layout.py** contains all Dash layout definitions, organizing the interface into modular, responsive sections. The layout utilizes the Dash Bootstrap Components to more easily build consistently styled apps with complex, responsive layouts.
- **graph_generator.py** includes all functions for generating line charts, heatmaps, and bar plots.
- **map_creator.py** defines and updates spatial visualizations using plotly.express.scatter_map library, enabling location-based filtering and drill-down functionality.

Dataframes that are created from scratch from the simulation data, such as departure and arrival times, are preprocessed into Parquet format using Pandas to reduce memory footprint and accelerate load times, especially for repeated visual queries.

CSV was chosen as the simulation output format due to its simplicity and compatibility with existing tools. To mitigate memory and performance limitations, large datasets are converted to columnar Parquet files using Pandas, enabling faster loading and selective querying. While no in-memory caching is used in the current version, future enhancements will explore query engines like DuckDB for on-disk filtering.

### 4.2 Dashboard Design

The dashboard interaction model reflects Shneiderman's information-seeking mantra [14], supporting 'overview first, zoom and filter, then details-on-demand' through the three-layer interface. Therefore, the front end, built using Dash by Plotly, comprises three presentation views, as described in the following.

**1) System Overview**
**KPI Summary Cards:** The top of the dashboard page in Fig. 2, displays global metrics such as transformer overload duration, first overload date, load factor, coincidence factor, total dissatisfaction events, average charging cost (DKK/kWh), average $CO_2$ emissions (kg/kWh), and DSO tariff revenue. Furthermore, the KPIs show the percentage difference for a reference scenario for easy comparison. The KPIs include a percentage difference relative to a baseline scenario when provided. Users can load a reference scenario via a file input, enabling dynamic comparison across key performance indicators. Differences are displayed as percentage changes while relevant graphs include an overlay of reference scenario, streamlining comparative scenario analysis.

**Spatial Map:** Shows agent locations color-coded by selected parameter (e.g., total electricity expenses, EV charging load, dissatisfaction count) as shown in Fig. 3. Clicking a point triggers a table with detailed information about the user. The map shows the statistics for the current selection with the sum, maximum, mean, and minimum for the system.

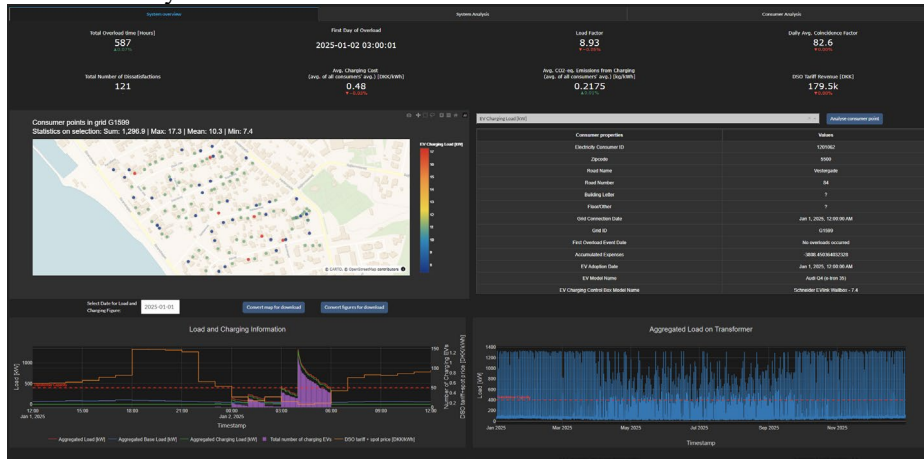

**Fig. 2.** The dashboard's system overview.

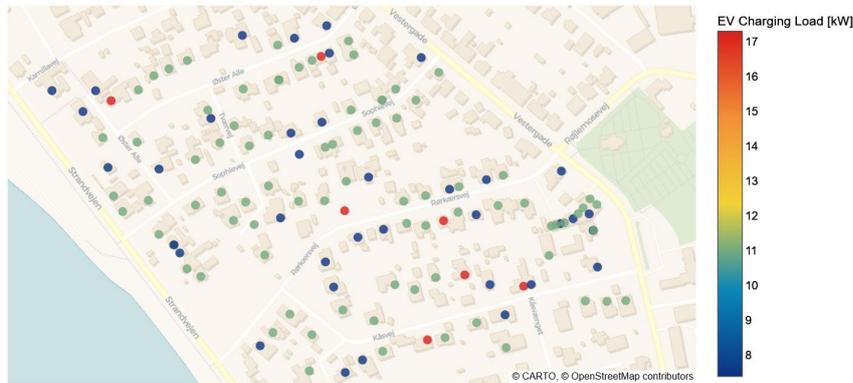

**Fig. 3.** Spatial map of the simulated system.

**Transformer Load Chart:** Aggregated load time series with capacity threshold indicated. The investigation of the system concerns the loading of the transformer and is therefore chosen to be located on the System Overview (bottom right in Fig. 2).

**Load and Charging Information Chart:** A chart (Fig. 4) showing in one graph the correlation between aggregated baseload, charging load, total load, electricity prices, the number of charging EVs, and the transformer capacity. This chart has a feature implemented enabling clicking on the bars for the number of charging EVs, which will show on the map the consumers charging at that specific point in time. This allows for easy access to consumer information at specific times in case some behavior requires deeper insights.

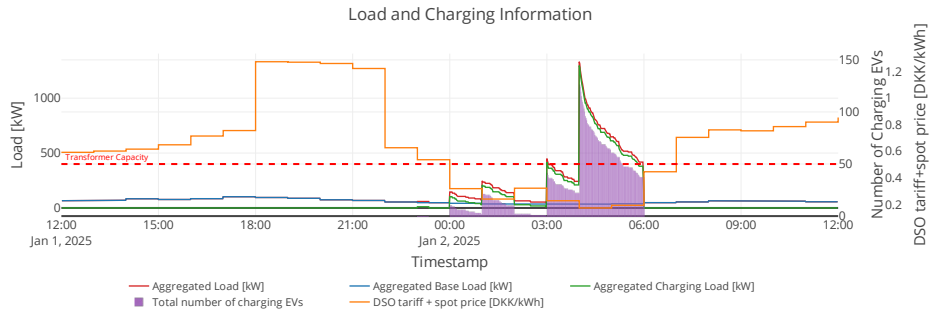

**Fig. 4.** Aggregated system load, EV charging load, baseload, electricity price, and number of charging vehicles over time. Clicking a bar displays consumer-level data on the spatial map.

## 2) System Analysis

**Time-Series Views:** Number of charging EVs; baseload; electricity price breakdown (spot, DSO tariff, total); $CO_2$ emissions as seen at the top of Fig. 5.

**Daily Heatmap:** EV charging load per user for a selected date, as shown in Fig. 6. The heatmap shows the charging load (color code) at time (x-axis) for all 126 users in the system (y-axis)

**Arrival/Departure Bar Chart:** Displays counts per time bin with hoverable agent IDs for anomaly tracing and is shown in Fig. 7.

**Overload Distribution:** Shows frequency of loading categories based on IEC 60076-7 [15] (Normal cyclic: 100–150%, Long-time emergency: 150–180%, Short-time emergency: 180–200%, Critical: >200%) and is shown in Fig. 8.

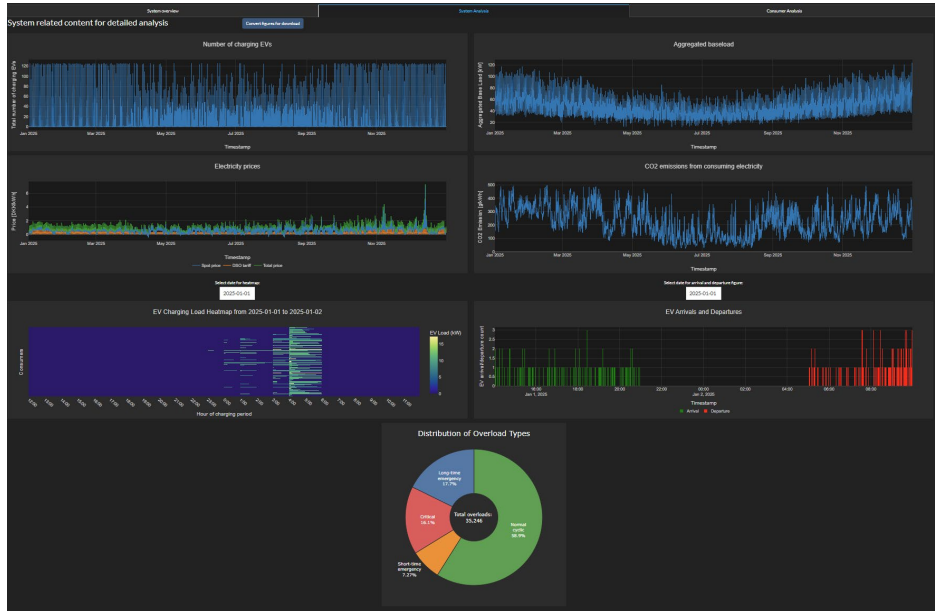

**Fig. 5.** The dashboard's System Analysis view page.

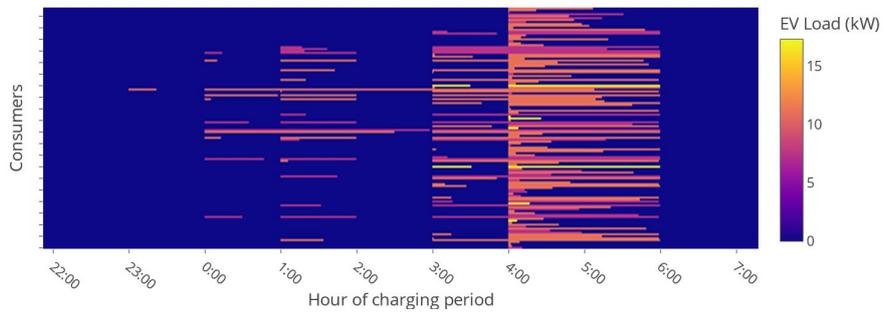

**Fig. 6.** Zoomed version of heatmap.

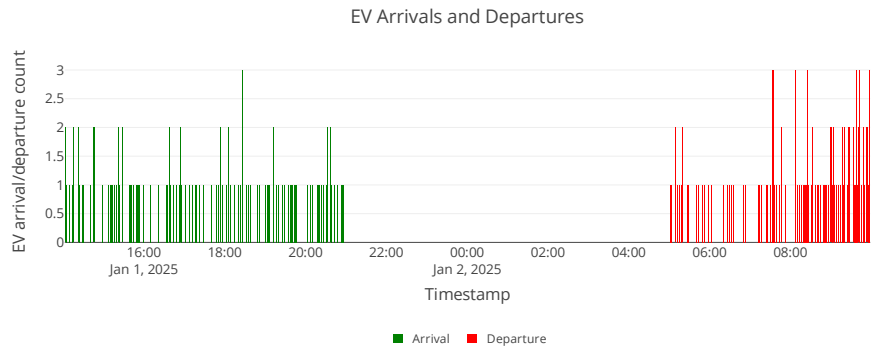

**Fig. 7.** EV arrivals and departure graph.

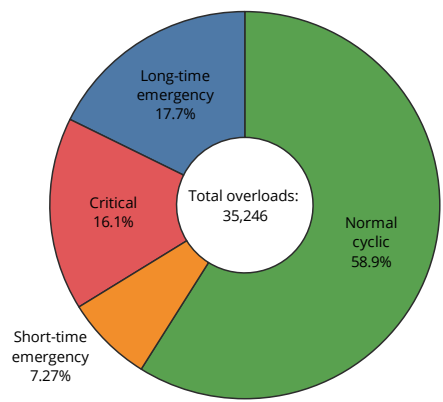

**Fig. 8.** Distribution of transformer overload types.

**3) Consumer Analysis**

**Agent Selection Input:** Text box for specific user ID, as shown in Fig. 9. From the System Overview page there is a button that automatically inserts the selected consumer (in the map) to the selection in Consumer Analysis page for fast access to information.

**Charging Load Step Chart:** User charging schedule with departure/arrival markers as shown in Fig. 10. This graph includes an optional reference overlay if selected from program initialization.

**Baseload, Driving Distance & SoC Charts:** Step charts showing the baseload consumption, daily driving distance of the consumer, and the SoC of the EV battery throughout the simulation. The SoC chart includes the feature highlighting days with insufficient charging level before departure, resulting in dissatisfaction. This is shown in Fig. 11, with the red cross added to each datapoint for the day with insufficient charge. The explanation for this unexpected behavior is found in a mismatch between the smart charging algorithm's handling of dates and the simulation dates when experiencing a change between summer and winter time.

All visualizations support pan, zoom, and have an export button for high-resolution SVG, with a "publication mode" to enlarge fonts and line styles for use in reports and articles. The publication mode has been used for all individual figures in this paper.

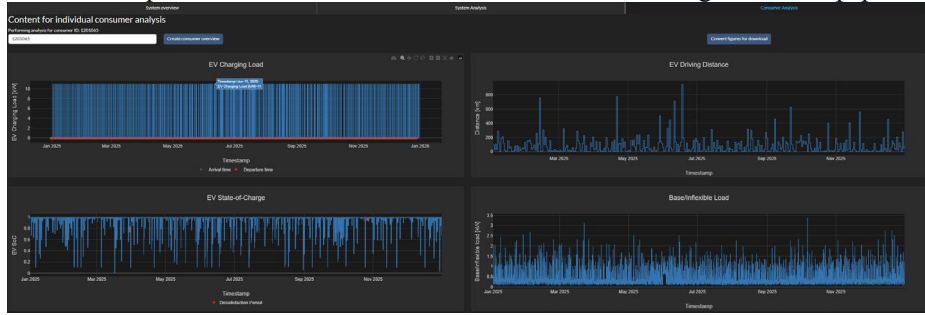

**Fig. 9.** The dashboards' Consumer Analysis view page.

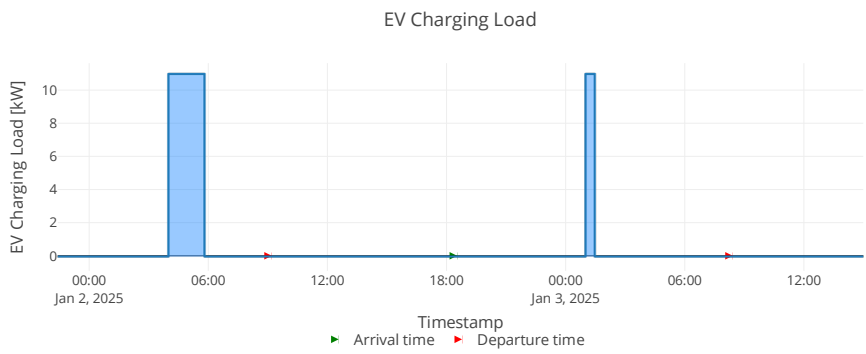

**Fig. 10.** Charging schedule for individual consumer analysis.

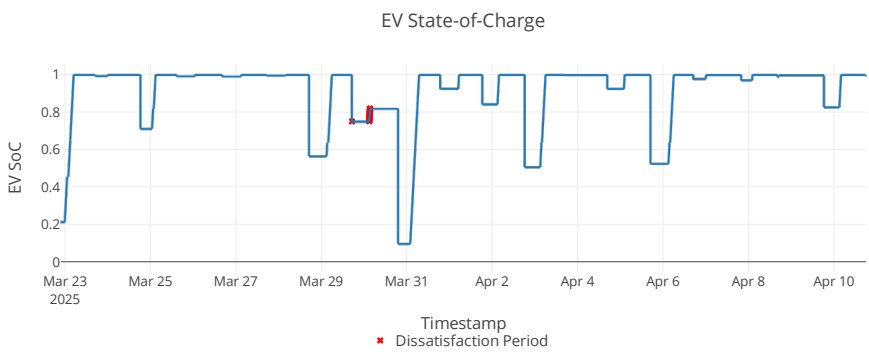

**Fig. 11.** SoC of individual EV batteries.

### 4.3    Case study highlights

• **Frequent overload events:** The correlation between the number of EVs at home, cheap electricity prices, and smart charging behavior results in daily overloads, with 16% of 587 hours of overload being critical (>200%).
• **Unwanted user dissatisfaction:** An analysis of user dissatisfaction events revealed that mismatches in daylight saving time transitions between the smart charging algorithm and simulation timestamps led to insufficient EV charging. This highlights the importance of consistent temporal alignment in simulation pipelines
• **Low utilization of existing capacity:** The low load factor indicates a very uneven distribution of the load, and the high coincidence factor confirms that many consumers locate their maximum load in the same period.

A deep understanding of the highlighted results above was rapidly identified and contextualized through synchronized dashboard views, which would have been difficult to detect with static scripts.

The dashboard supports both researchers and DSOs in analyzing scenario performance and investigating unexpected system behaviors. While the tool itself offers a shared set of indicators and interactive views, its application depends on the simulated scenario and user objective. For example, a DSO might explore grid impact under different tariff schemes, while a researcher could examine behavioral responses or test control strategies. In both cases, the dashboard facilitates decision support by making complex results more accessible.

## 5    Discussion

The proposed dashboard framework facilitates comprehensive analysis of complex outputs from MABS, enabling both system-level pattern recognition and detailed investigation of agent-specific anomalies. By prioritizing the identification and interpretation of emergent behaviors—such as localized transformer overloads or user-level dissatisfaction events—the tool provides researchers and distribution system operators with a practical means to derive actionable insights. This is particularly valuable when evaluating the effectiveness of strategies like time-of-use tariffs or controlled charging schemes.

A key strength of the framework lies in its integration of high-resolution interactive visualizations with a domain-specific data processing backend. Users can explore large volumes of stochastic simulation data through synchronized views that support anomaly detection and root-cause analysis without extensive manual preprocessing.

The dashboard supports both researchers and DSOs in analyzing scenario performance and investigating unexpected system behaviors. While it offers a shared interface and consistent indicators, its application varies by user objective. For instance, DSOs may focus on infrastructure stress under different tariff schemes, while researchers might examine behavioral adaptations or validate control strategies. This flexibility enhances the tool's value across both operational planning and academic exploration.

However, the framework does have limitations regarding data volume. In its current implementation, the dashboard has been tested with datasets comprising 1-minute resolution measurements over a one-year simulation period, producing CSV files of approximately 8 GB per scenario. Despite the substantial size, the dashboard performs reliably after the initial loading phase, including when comparing two full scenarios (e.g., baseline and test case).

As simulation scales increase—whether in temporal resolution, spatial granularity, or model complexity—future work should consider integrating efficient, queryable storage solutions such as DuckDB or similar in-process analytical databases. These would enable on-demand filtering and aggregation directly on disk, thereby reducing memory overhead and improving responsiveness for larger datasets.

Overall, the framework offers a scalable foundation for exploratory analysis of complex MABS outputs, and its modular design supports future enhancements both in terms of data volume handling and extended visualization capabilities. For example, in the case study, the modular structure allows for simple extension to other distributed energy resources (e.g., home batteries, heat pumps), further increasing the dashboard's relevance.

While formal usability evaluation is pending, early informal feedback from internal stakeholders highlighted the value of the dashboard's layered navigation and drill-down capabilities. A structured validation process is planned, including direct involvement of external stakeholders, such as DSOs, to assess usability, interpretability, and operational impact. Future evaluations will follow established protocols such as heuristic analysis and the System Usability Scale, supporting iterative improvements and more robust validation of the tool's decision-support effectiveness.

## 6    Conclusion

This paper introduced an interactive dashboard framework tailored for analyzing outputs from multi-agent-based simulations (MABS) of electric vehicle (EV) home charging ecosystems. The tool enables researchers, distribution system operators (DSOs), and energy planners to explore complex, high-frequency simulation data across multiple analytical levels—from system-wide metrics to individual user behavior—using coordinated visualizations and agent-level drill-downs.

By focusing on emergent phenomena rather than raw computational performance, the framework supports the identification and contextual explanation of unexpected system behaviors, such as clustered transformer overloads or spikes in consumer dissatisfaction. This capability is critical in domains where stochastic agent interactions and temporal flexibility lead to highly dynamic outcomes that are difficult to trace using traditional static plots or aggregated indicators.

The architecture leverages a modular Python backend and Dash by Plotly for front-end interactivity, with efficient data preprocessing strategies to manage year-long, minute-resolution datasets. Through a detailed case study, the dashboard was shown to rapidly uncover system inefficiencies, behavioral anomalies, and potential mis-matches in temporal logic—insights that would be challenging to obtain through manual script-based analysis alone.

While current limitations include data scalability and visualization support for non-EV distributed energy resources, the framework is designed for extensibility. Future work will incorporate additional distributed energy resource types such as photovoltaic systems, household batteries, and heat pumps, already present in the simulation engine but not yet visualized. Furthermore, integration with query-optimized data handling (e.g., DuckDB or Apache Arrow) is planned to support the exploration of even larger and more complex simulation scenarios.

Overall, the dashboard lowers the barrier to high-resolution exploratory analysis of MABS outputs and serves as a practical decision-support tool for evaluating the effectiveness of smart grid interventions in complex, agent-driven systems.

## Acknowledgement

This paper is part of the project titled "Automated Data and Machine Learning Pipeline for Cost-Effective Energy Demand Forecasting in Sector Coupling" (jr. Nr. RF-23-0039; Erhvervsfyrtårn Syd Fase 2), The European Regional Development Fund.

## Appendix

**Table 1.** Dashboard tool feature comparison between popular tools.

| Feature | Dash by Plotly [16] | Power BI [17] | Tableau [18] | Streamlit [19] | Shiny [20] |
|---|---|---|---|---|---|
| Languages | Python, R, Julia | Power Query (M), DAX, supports Python/R scripts | Drag & drop UI, VizQL; R/Python via extensions | Python | R (original), Python (new) |
| Visualization Engine | Plotly.js (HTML/SVG/WebGL); deeply integrated | Microsoft proprietary engine (TypeScript/HTML) | VizQL + Hyper engine | Vega-Lite, Plotly, Matplotlib, etc. | ggplot2, Plotly, htmlwidgets, Vega, Leaflet |
| Interactivity | High: callback-driven, multi-page, fully programmable | High: dashboard filters, drill-throughs | High: filters, parameters, tooltips | Medium: script reruns on input | High: reactive model, dynamic UI |
| Scientific Use Fit | Excellent: MABS, time series, system modeling | Limited: Business KPIs; less flexible for research | Moderate: supports R/Python, but GUI-bound | Good: fast for prototyping, but limited structure | Excellent: highly used in academic/statistical apps |
| Data Volume Handling | Good: needs pre-processing/parquet for large data | Very high (with Premium/DirectQuery) | High (Hyper extracts or live DB connections) | Moderate: in-memory; backend scaling needed | Moderate: in-memory limits; scale with DBs |
| Deployment | Self-hosted (Flask), Dash Enterprise (proprietary) | Microsoft eco-system: Desktop, Cloud, Report Server | Tableau Server, Cloud, Public | Local, Docker, Streamlit Cloud | Shiny Server (open-source), RStudio Connect (pro) |
| Licensing | Open-source SDK (MIT); proprietary hosting optional | Proprietary (no open-source components) | Proprietary (owned by Salesforce) | Fully open-source (Apache 2.0); cloud service optional | Open-source (GPL/MIT); commercial hosting optional |
| Learning Curve | Moderate: Python + callback logic | Low–moderate: GUI-based, Excel-like | Low–moderate: visual modeling, Tableau calc lang | Very low: basic Python scripting | Moderate: easy to start, advanced logic requires understanding reactivity |
| Customization & Extensibility | Very high: HTML/CSS, JS, React components | Limited to prebuilt/custom visuals | Moderate: themes, extensions API | Moderate: layout primitives, custom components | Very high: full HTML/CSS/JS control |